\documentclass[preprint]{aastex}
\usepackage{times}
\usepackage{natbib}
\usepackage{graphicx}

\newcommand{\eb}[1]{\begin{equation}\label{eq:#1}}
\newcommand{\en}{\end{equation}}
\newcommand{\eqa}[1]{\begin{eqnarray}\label{eq:#1}}
\newcommand{\eqn}{\end{eqnarray}}

\newcommand{\Fig}[1]{{Figure~\ref{fig:#1}}}

\shortauthors{Stein and Nordlund}

\begin{document}

\title{On the Formation of Active Regions}
\author{Robert F. Stein}
\affil{Department of Physics \& Astronomy,
 Michigan State University, East Lansing, MI 48824, USA}
\email{stein@pa.msu.edu}
\author{{\AA}ke Nordlund}
\affil{Niels Bohr Institute, University of Copenhagen, DK-2100 Copenhagen, DK}
\email{aake@nbi.dk}

\begin{abstract}
Magneto-convection can produce an active region without an initial
coherent flux tube.  A simulation was performed where uniform,
untwisted, horizontal magnetic field of 1 kG strenght was advected
into the bottom of a computational domain 48 Mm wide by 20 Mm deep.
The up and down convective motions produce a hierarchy of magnetic
loops with a wide range of scales, with smaller loops riding ``piggy
back" in a serpentine fashion on larger loops.  When a large loop
approaches the surface it produces an small active region with a
compact leading spot and more diffuse following spots.
\end{abstract}


\section{Introduction}
The standard paradigm is that active regions form when a coherent
flux tube from deep in the convection zone reaches the surface, typically
modeled using the thin flux tube approximation
\citep{Parker55,Fan93,Moreno-Insertis94,Caligari95,Fan09,Weber11}.
\citet{Cheung07,Martinez-Sykora08,Weber11,Fang12} have shown how
important the actual convective motions are to the rise of magnetic
flux.  Recent simulations of active region formation have started
from a coherent semi-torus of magnetic field placed in the surface
layers of a model solar convection zone \citep{Cheung10}.  Our
simulations show that such a coherent structure is not necessary
for the formation of an active region.  The action of magneto-convection
itself produces rising flux tubes, which when they reach the surface
can produce an active region.

\section{The Simulation}
We use the {\sc stagger code} \citep{Beeck12} to simulate
magneto-convection in a domain 48 Mm wide by 20 Mm deep.  This depth
is 10\% of the geometric depth of the solar convection zone, but
contains two thirds of its pressure scale heights.  Horizontal
directions are periodic and the vertical boundaries are open with
plasma and magnetic field moving through them.  Because of the large
domain dimensions the coriolis force from the solar rotation begins
to have some effect, so we include f-plane rotation at a lattitude
30 deg north.  The initial state was a snapshot of hydrodyanamic
convection which had been relaxed for several turnover times (about
2 days at 20 Mm depth).  We then started advecting minimally
structured, uniform, untwisted, horizontal magnetic field into the
domain by inflows at the bottom.  The initial field strength was
200 G (weak enough to have little dynamic effect), which was slowly
increased with a 5 hour e-folding time until it reached 1 kG strength
and thereafter held constant.  The incoming field is at 30 deg to
the east-west axis.  The active region is not affected by the 
horizontal boundaries: first, because its size is 25 Mm inside a
48 Mm wide box, second, because it is spreading very slowly with
time and, third, because the top boundary is close by (at the
temperature minimum) so the magnetic field does not have room to
spread out much in the atmosphere and interact with the mirror
active regions produced by the horizontal periodicity.

\section{Formation of an active region}

\begin{figure}[!htb]
 \centerline{\includegraphics[width=0.25\textwidth]{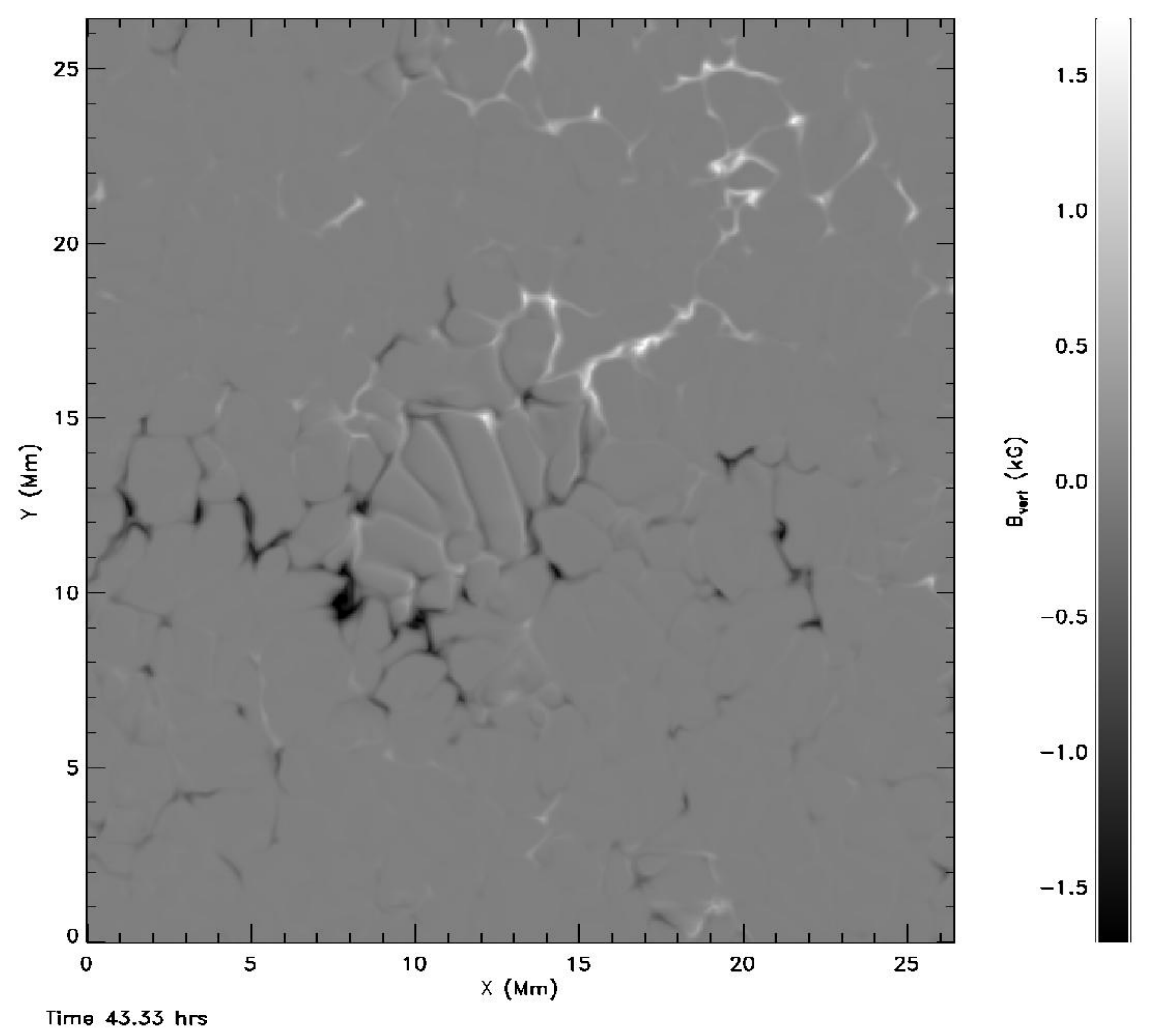}
 \includegraphics[width=0.25\textwidth]{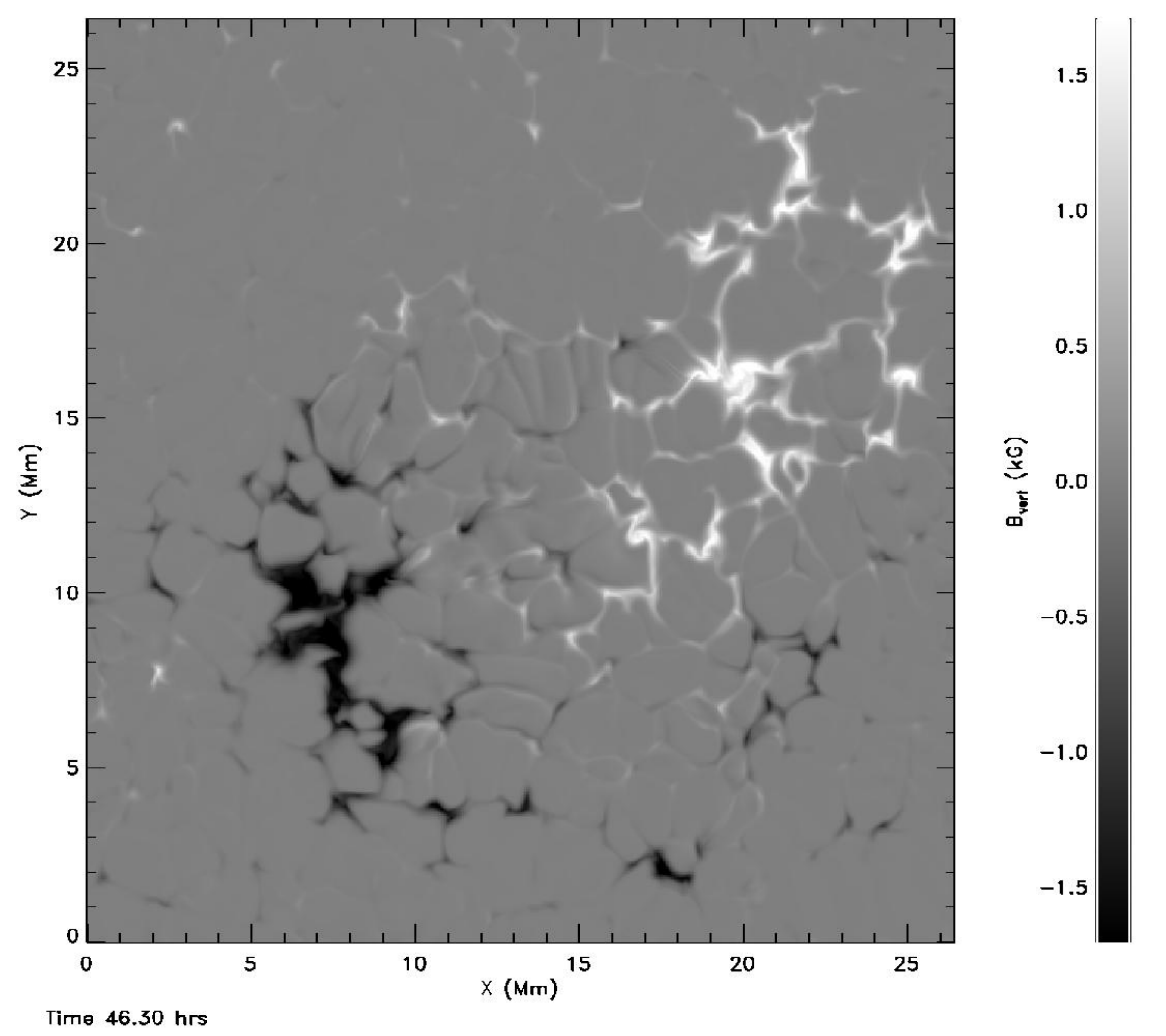}
 \includegraphics[width=0.25\textwidth]{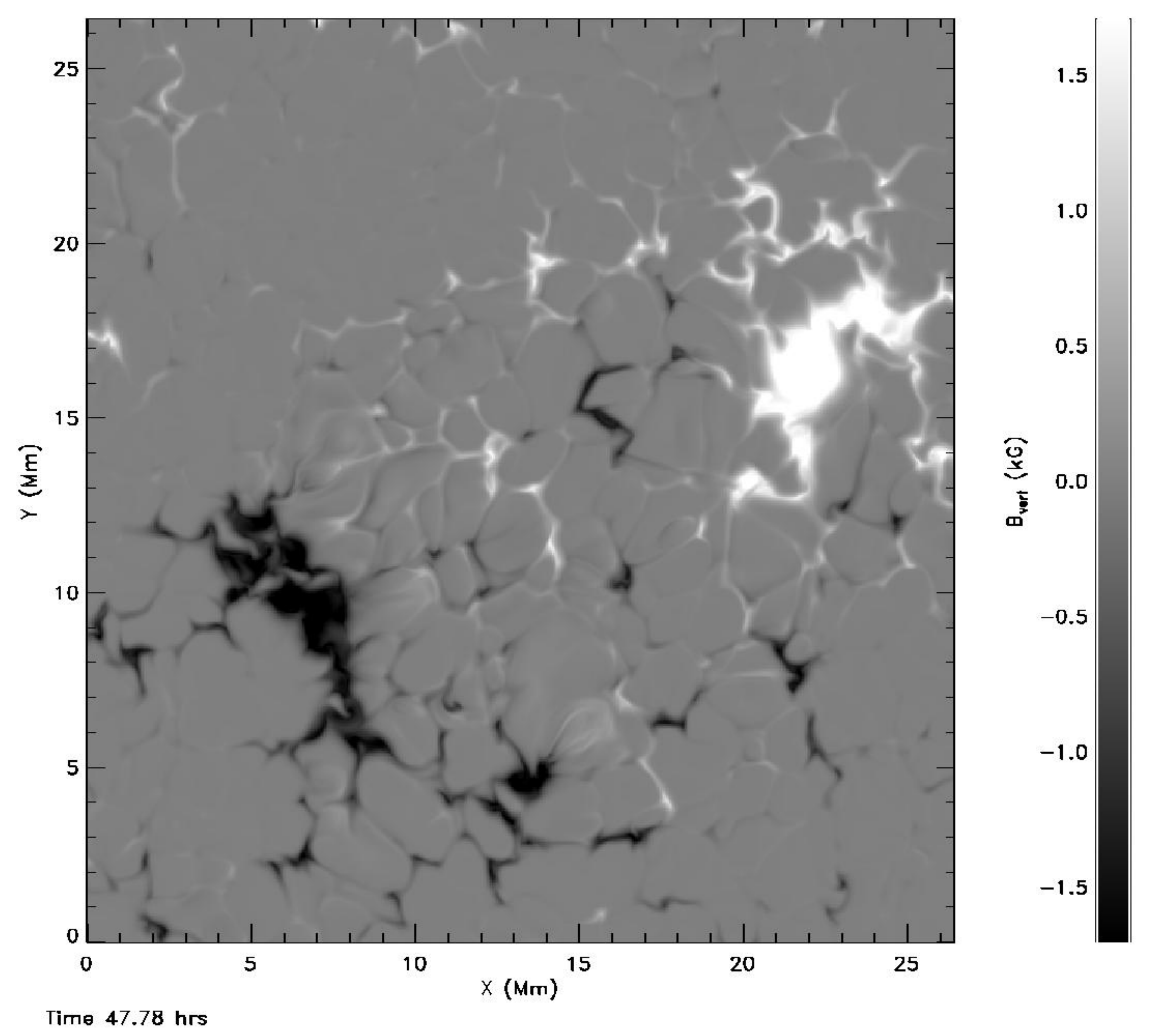}
 \includegraphics[width=0.25\textwidth]{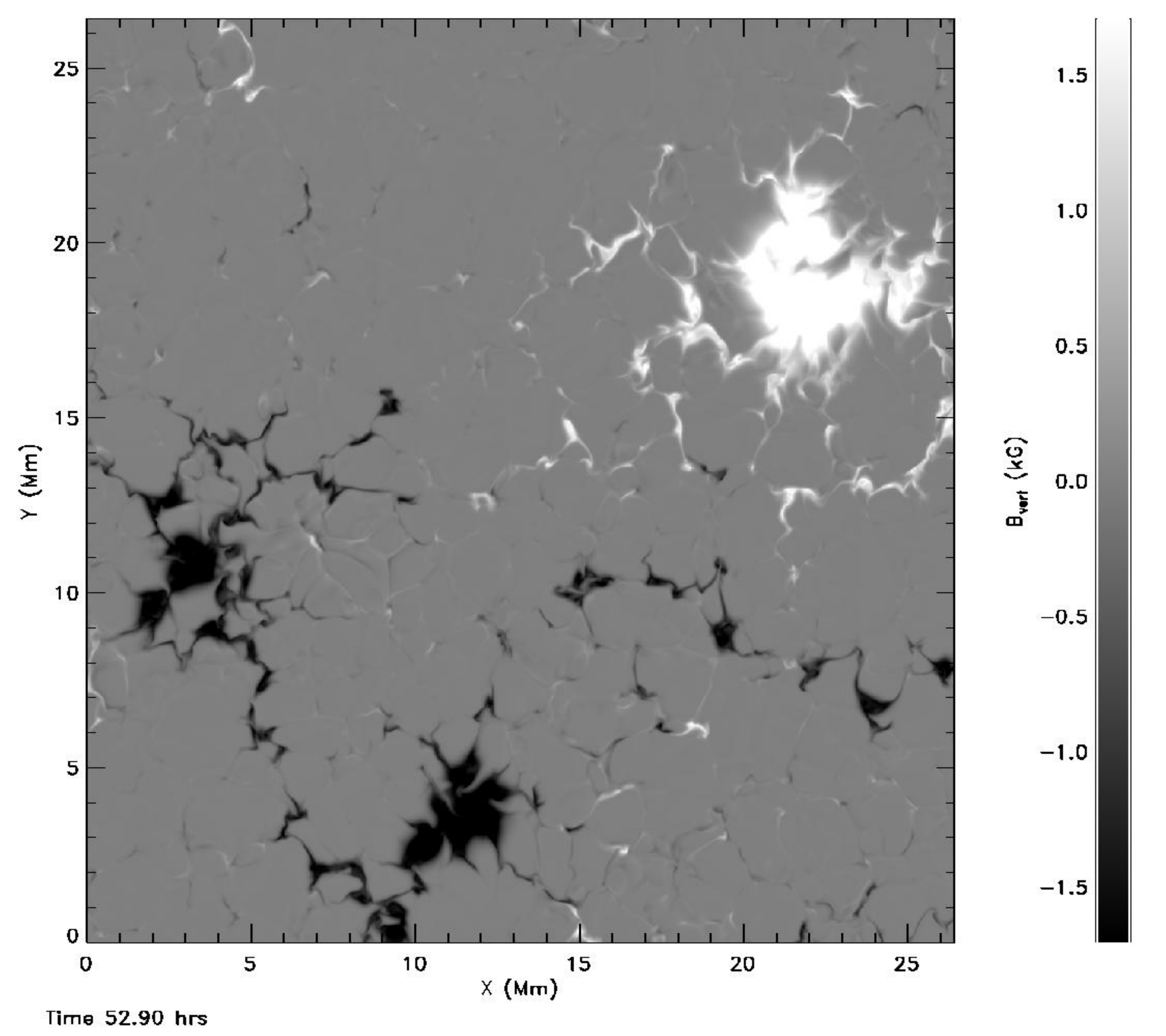}}
\caption{Vertical magnetic field.  The images are clipped at $\pm$
1.7 kG, but the actual range is $\pm$ 3 kG.  The x-axis is the East-West
axis.  The field entered in the upflows at the bottom at an angle
30$^o$ to the x-axis.  The AR approximately maintains this orientation.
The separation of the pores is approximately the size of the 
supergranule scale convective cells near the bottom of the domain.
}
\label{fig:ARform1}
\end{figure}

\begin{figure}[!htb]
 \centerline{\includegraphics[width=0.25\textwidth]{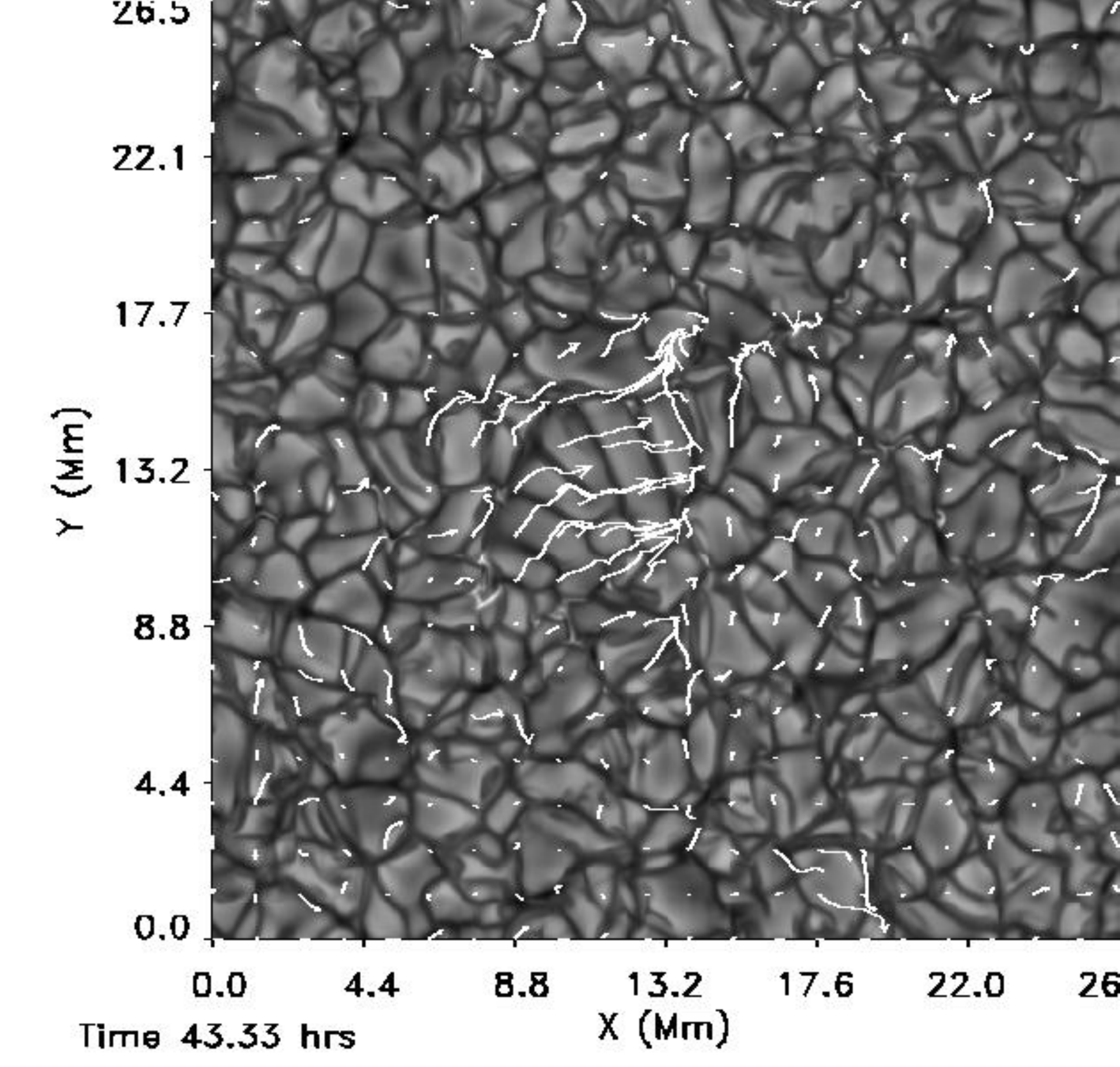}
 \includegraphics[width=0.25\textwidth]{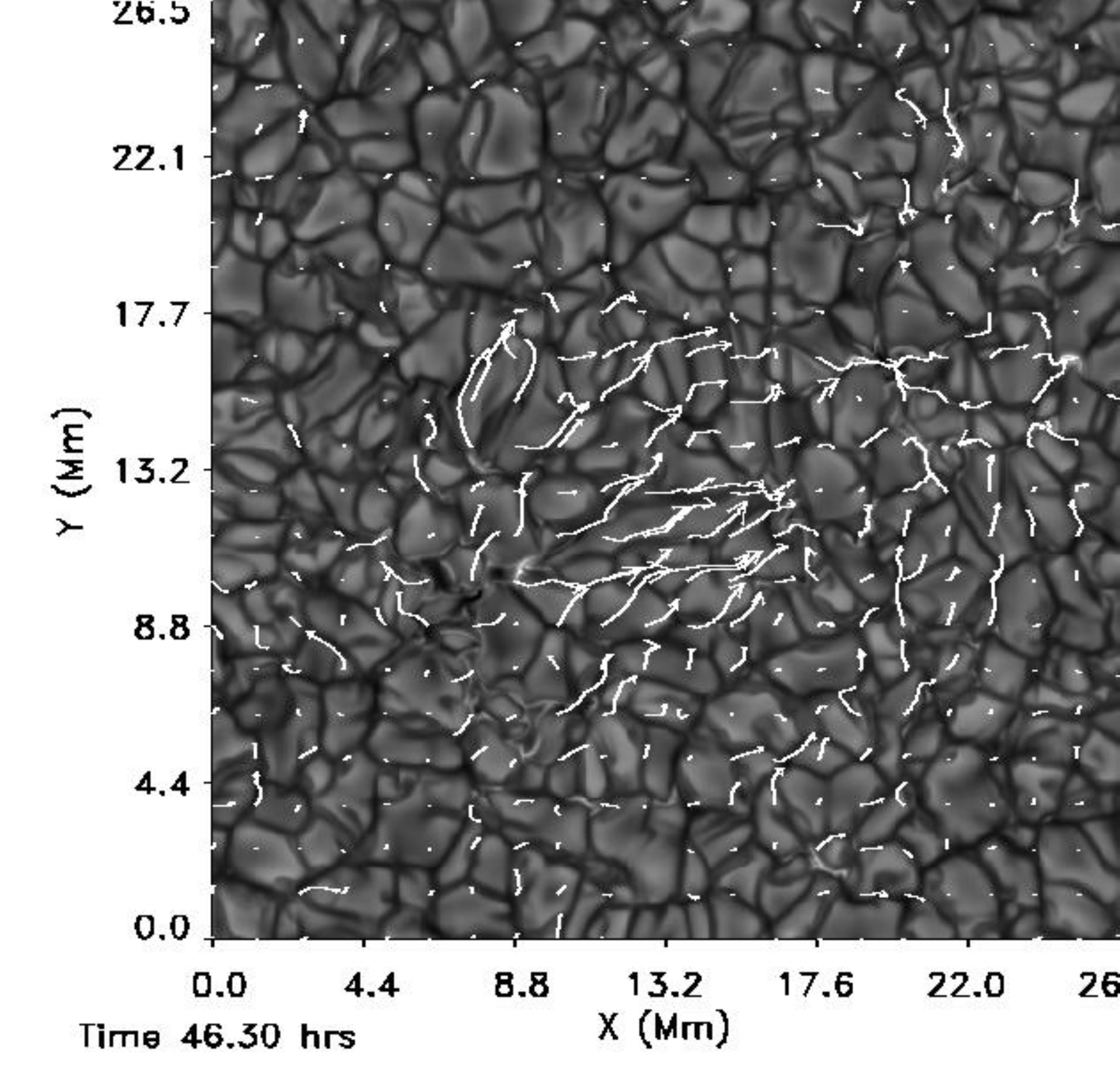}
 \includegraphics[width=0.25\textwidth]{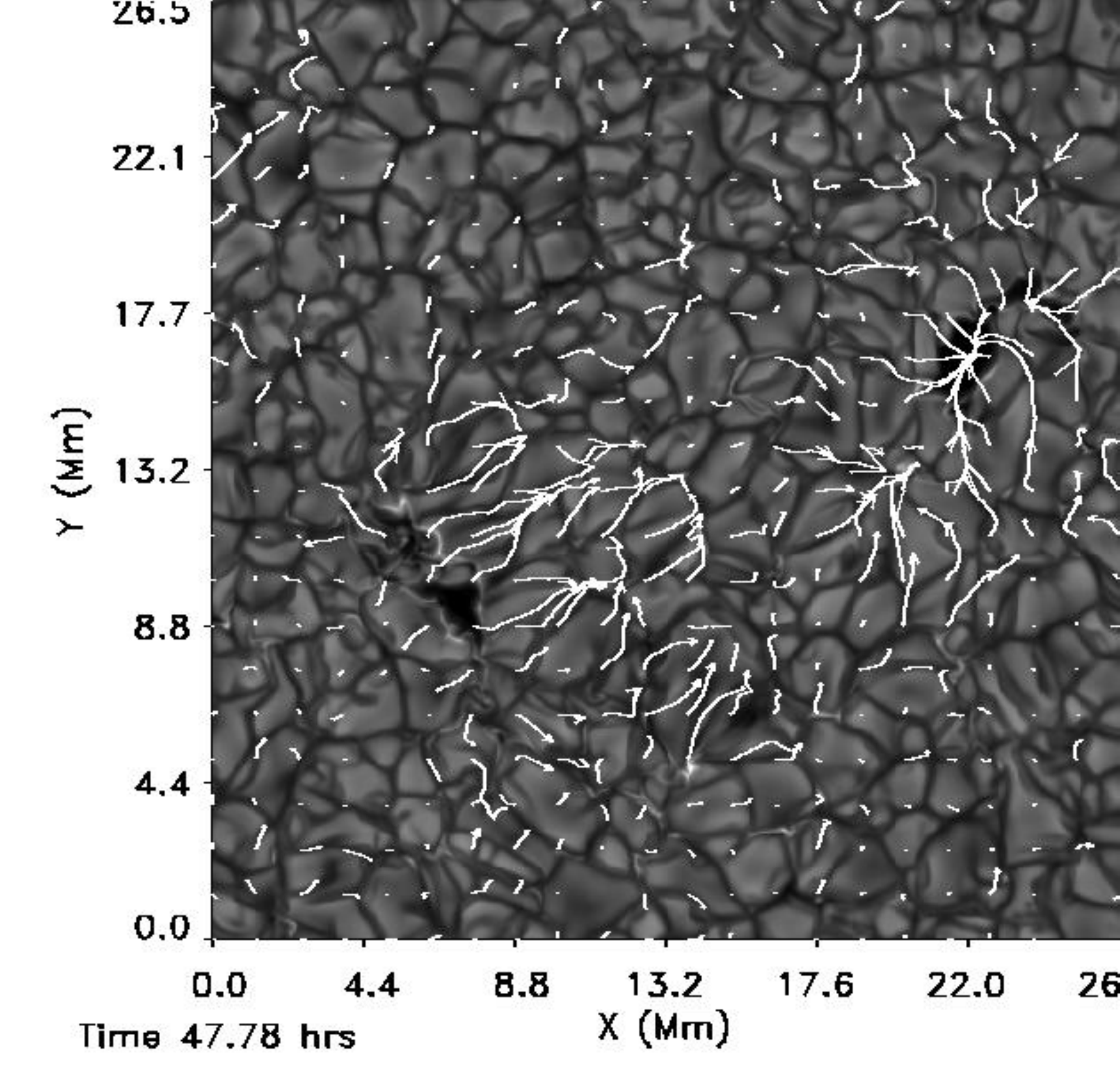}
 \includegraphics[width=0.25\textwidth]{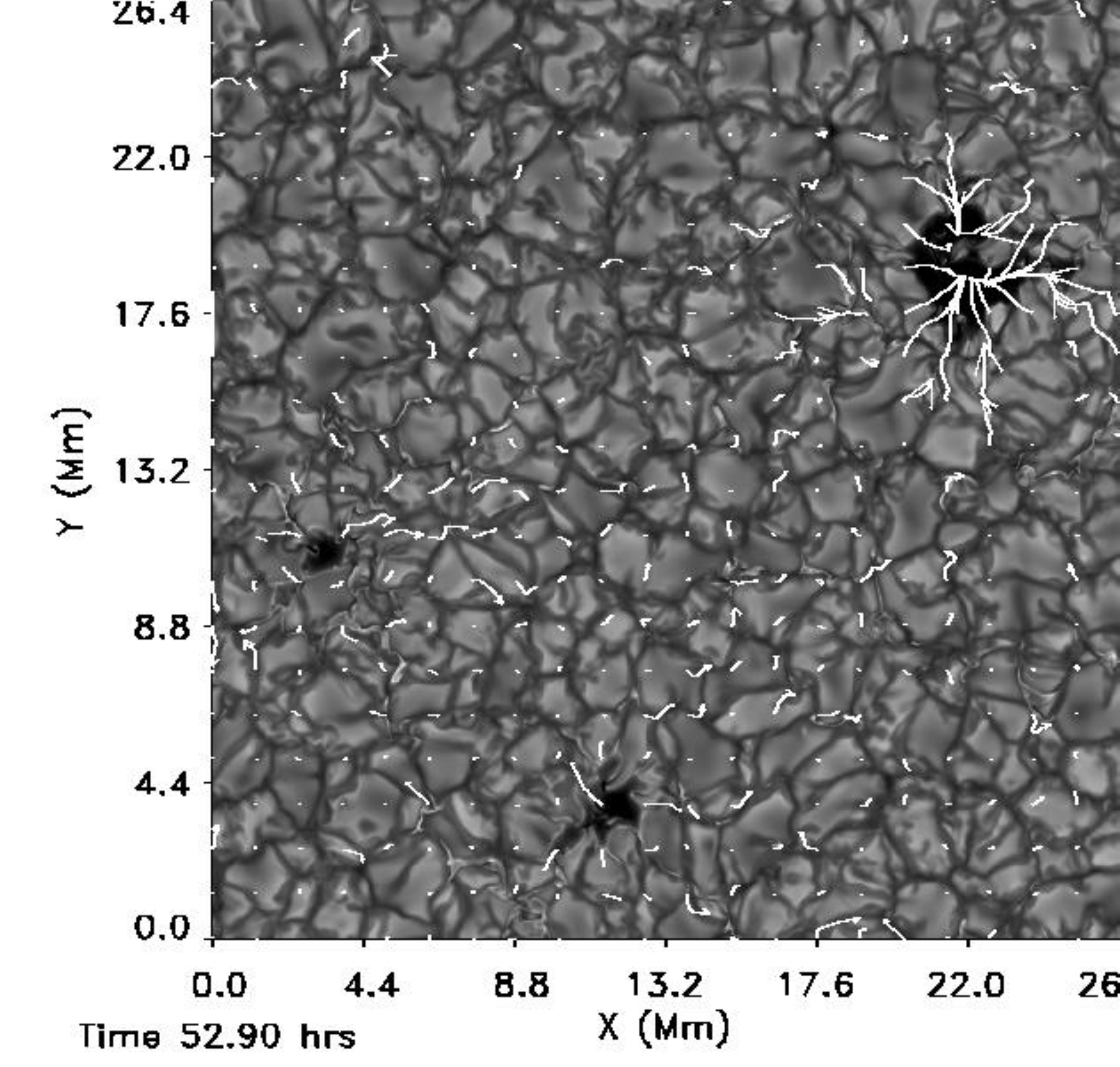}}
\caption{Continuum intensity image with horizontal magnetic field
vectors superimposed.  The images are clipped at $2.3 > I/\left<I\right> >
0.5$.   The actual range is [0.2,2.5].  In the initial emergence the
granules are elongated transverse to the honrizontal field.  Thereafter
the granules appear elongated along the magnetic field direction.
}
\label{fig:ARform2}
\end{figure}

\begin{figure}[!htb]
 \centerline{\includegraphics[width=0.25\textwidth]{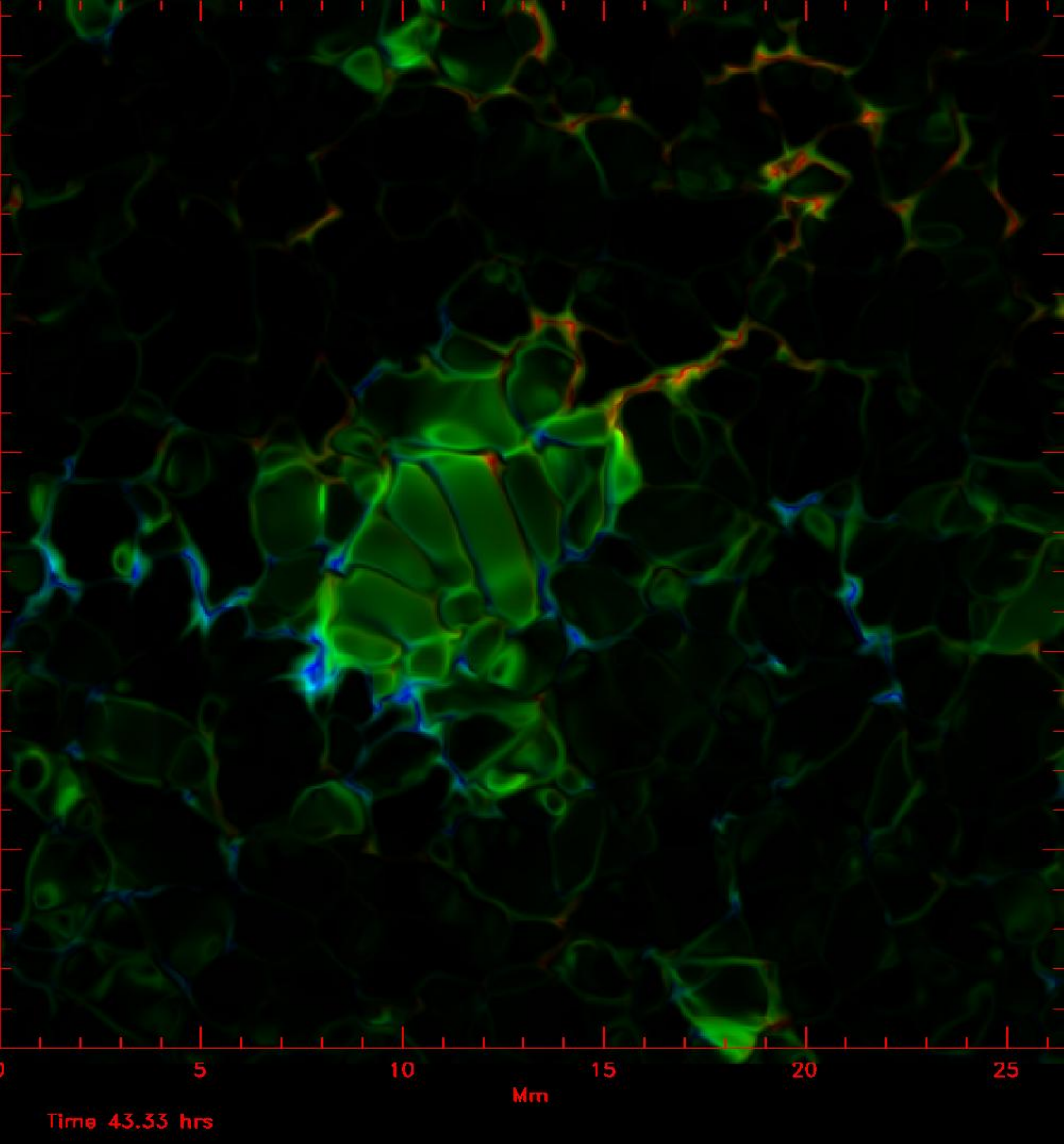}
 \includegraphics[width=0.25\textwidth]{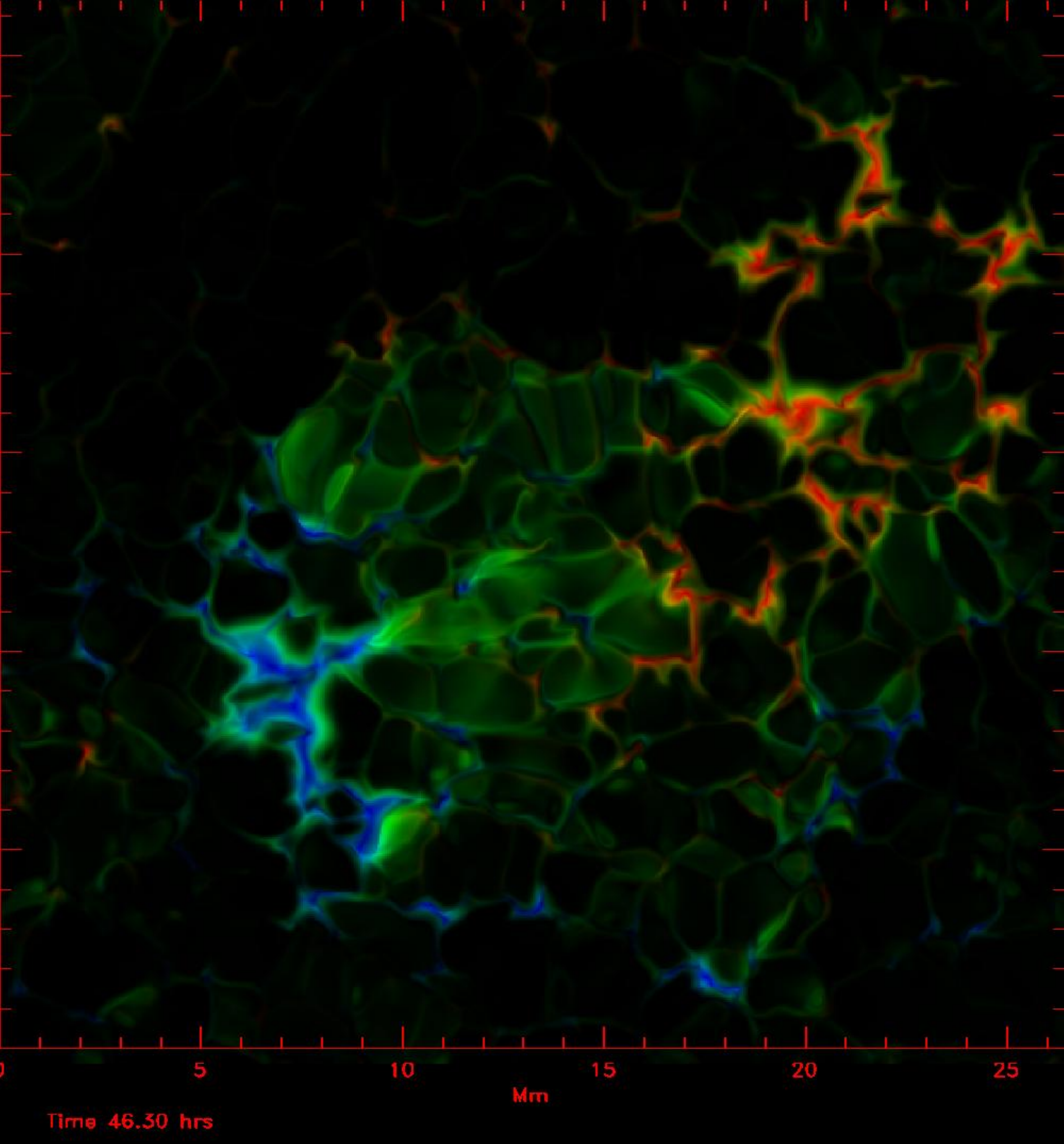}
 \includegraphics[width=0.25\textwidth]{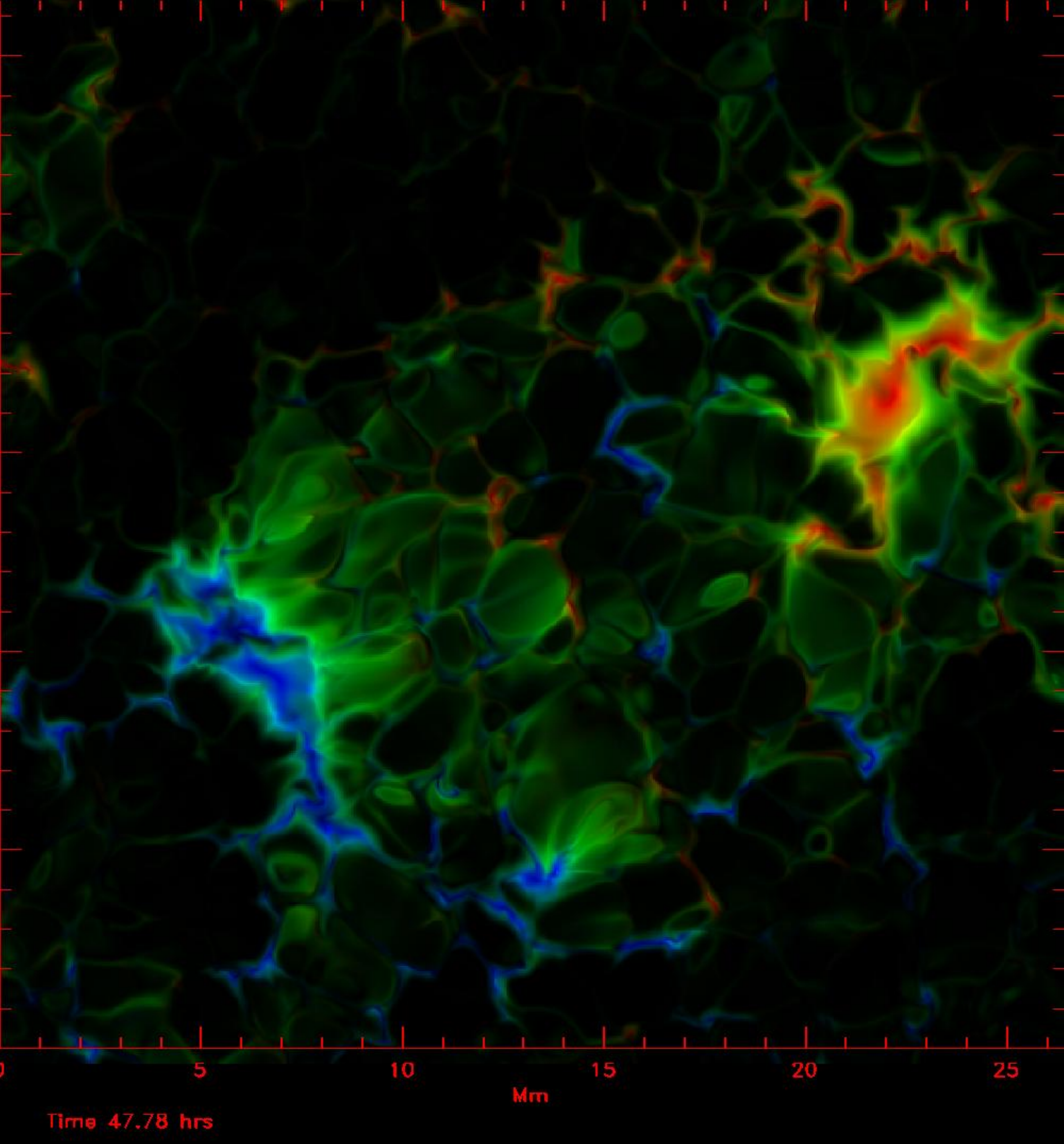}
 \includegraphics[width=0.25\textwidth]{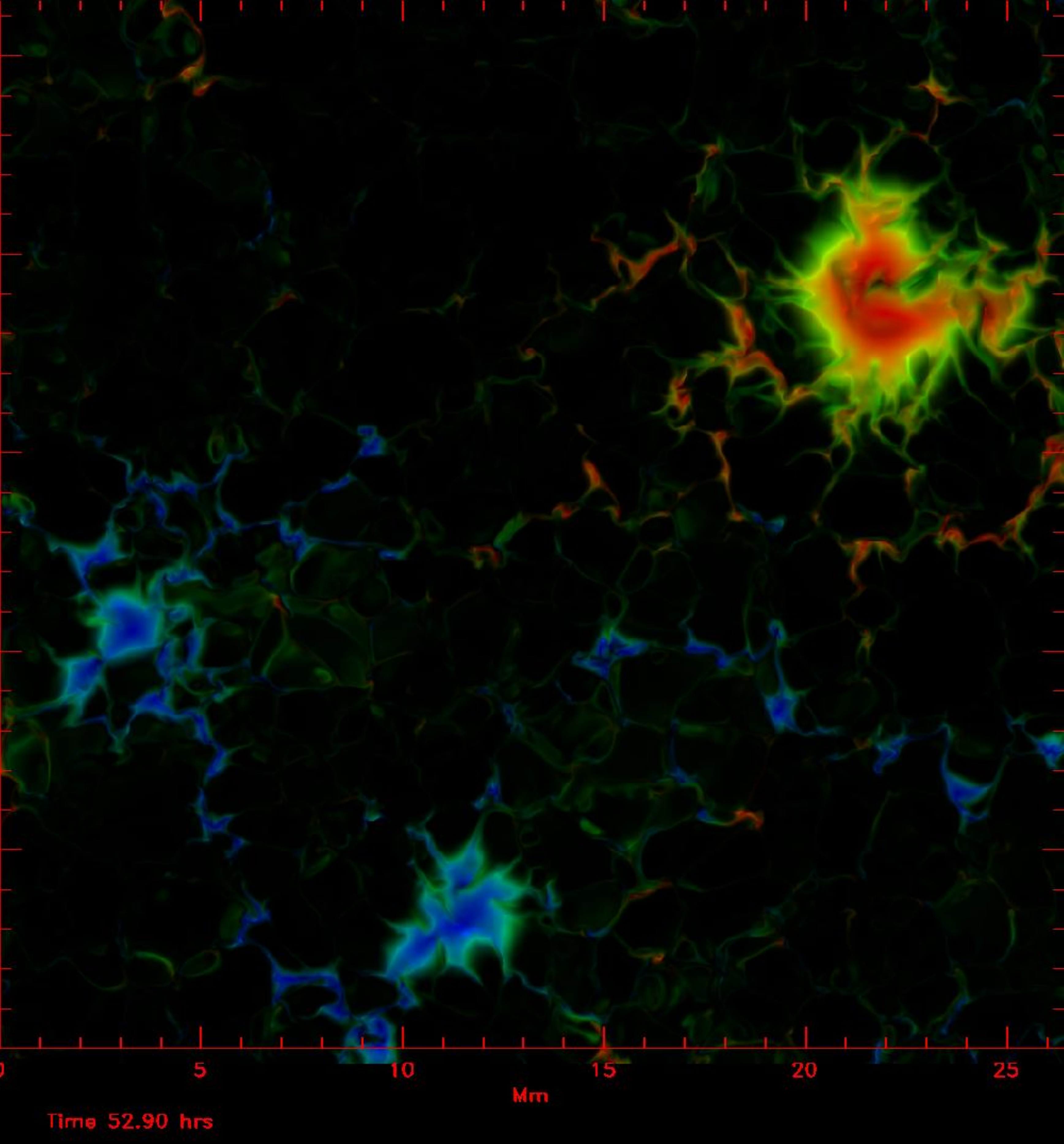}}
\caption{Image of horizontal (green) and vertical (blue and red)
magnetic field.  Magnetic field first emerges horizontally over
granules followed by the appearance of vertical field at the granule
edges.  The orizontal field is quickly swept into the intergranular
lanes.
}
\label{fig:ARform3}
\end{figure}

As the magnetic field enters the computational domain, convective
upflows immeadiately begin to drag portions toward the surface, while
convective downflows immeadiately pull other portions down, thus creating
magnetic loops.  
Magnetic field begins to reach the surface after about 20 hours.
Smaller scale convective patterns nearer the surface produce
smaller scale crenulations in the magnetic loops, giving them
a serpentine structure with small loops riding ``piggy-back" on
larger ones.  The field first emerges as small bipoles
with mixed polarity over a confined region of 25 Mm square.
At about 40 hours, a large flux bundle begins to emerge with rate
of 2-4$\times$10$^{20}$ Mx/hr  
(\Fig{ARform1}).  The area of major flux emergence is confined by 
the convective up- and down-flows and does not fill the entire 
domain.  Flux emergence in the remainder of the domain is
much smaller.  The field first emerges horizontally over granules
and then vertical field appears in the intergranule lanes at the
ends of the horizontal field.  The vertical legs of the bipoles quickly
separate.  Initially, granules become elongated transverse to the
magnetic field.  In later emergence events the granules are elongated
parallel to the horizontal field (Figures~\ref{fig:ARform2} and \ref{fig:ARform3}).
Opposite polarity field migrates to unipolar clusters, sometimes
colliding and cancelling in the process (\Fig{ARform1}).  The horizontal
components of the emerging field pass through the upper boundary,
leaving behind their nearly vertical legs.  Individual flux
concentrations collect.  The leading spot is more compact and
rotating and the follow spots more diffuse and non-rotating
(\Fig{ARform1}).  The unsigned vertical flux in the active region is
a little more than $10^{21}$ Mx and a similar amount is in the
remainder of the domain.  This active region has not formed penumbra.
As found by \citet{Rempel12} and upper boundary condition forcing
a more horizontal field at the surface is needed to form penumbra.
The emerging flux and active region here covers a region of
supergranule scale, similar to the sizes of the convective cells
at the bottom of the simulation domain.  The orientation of the
active region is approximately the 30 deg orientation of inflow
field at the bottom.  The magnetic field lines that emerge are an
undulation on a larger, subsurface magnetic structure spanning most
of the domain.  Eventually, the rate of new flux emergence slows
and the space between the spots has little horizontal field
(\Fig{ARform3}).

\section{Summary and Conclusions}

These simulations show that magneto-convection itself can produce
the flux tubes that give rise to active regions.  The
action of up and downflows on an initial supergranule size patch
of horizontal field can keep the major portion of the magnetic flux
confined to emerge at the surface in a similar supergranule size
region.  Interesting work for the future will be to investigate the
evolution of larger scale magnetic structures, such as those in the
flux emergence simulations of \citet{Abbett00} and \citet{Fan08},
as they pass through the upper convection zone.

Movies of the active region formation are available on 
http:steinr.pa.msu.edu/$\sim$bob/research.html\#AR.

\acknowledgements

The calculations were performed on the Pleiades cluster of the NASA Advanced Supercomputing Division.
RFS is supported by NASA grants NNX12AH49G and NNX08AH44G and NSF grant AGS-1141921.
This support is greatly appreciated.

\bibliographystyle{apjbib}

\end{document}